\documentclass[%
 reprint,
 amsmath,amssymb,
 aps, 
 pre,
 showkeys
]{revtex4-1}

\usepackage{graphicx}
\usepackage{dcolumn}
\usepackage{bm}

\usepackage{my-style}
\graphicspath{{figs/}}

\begin{document}

\preprint{APS/123-QED}

\title{Susceptibility Propagation by Using Diagonal Consistency}

\author{Muneki Yasuda}  
\affiliation{%
 Graduate School of Science and Engineering, Yamagata University
}%
\author{Kazuyuki Tanaka}%
\affiliation{%
 Graduate School of Information Sciences, Tohoku University
}%

\begin{abstract}
A susceptibility propagation that is constructed by combining a belief propagation and a linear response method is used for approximate computation for Markov random fields. 
Herein, we formulate a new, improved susceptibility propagation by using the concept of a diagonal matching method that is based on mean-field approaches to inverse Ising problems. 
The proposed susceptibility propagation is robust for various network structures, 
and it is reduced to the ordinary susceptibility propagation and to the adaptive Thouless-Anderson-Palmer equation in special cases.
\end{abstract}

\pacs{Valid PACS appear here}
\maketitle


\section{Introduction} \label{sec:intro}

There is an increased demand for techniques that can be used to evaluate local statistical quantities such as local magnetizations and local susceptibilities (covariances) 
of spin systems with finite sizes in statistical physics as well as in other scientific fields. 
This is because the use and application of Markov random fields and probabilistic operations is widely prevalent in many areas, 
particularly in the field of information sciences~\cite{Opper&Saad2001, Mezard&Montanari2009}.

Linear response methods have been used to approximately evaluate pair correlations between non-nearest pairs by combining an advanced mean-field method, 
particularly the cluster variation method, with the linear response theory~\cite{Sanchez1982, Morita1990}. 
In the current decade, a suitable technique called susceptibility propagation (SusP) (also called a variational linear response in the field of information sciences), 
has been developed to compute the local susceptibilities of Markov random fields~\cite{K.Tanaka2003, Welling&Teh2003, Welling&Teh2004, Mezard&Mora2009}. 
In general, SusPs are constructed by combining belief propagation algorithms and linear response methods.
Belief propagation algorithms are one of the most popular message-passing type of algorithms 
that is used to approximately compute local magnetizations of Markov random fields~\cite{Pearl1988}, 
and they are equivalent to Bethe approximations \cite{Bethe1935} used in statistical physics~\cite{Kabashima&Saad1998, GBP2001}.

It is known that SusPs can be used to compute local susceptibilities with a high degree of accuracy because linear response methods 
partially restore the effects of loops of networks that are neglected in the Bethe approximations~\cite{Yasuda&Tanaka2007}. 
SusPs have the following inconsistency due to approximation.
In Ising models, the variances, $\ave{S_i^2}$, are trivially one because $S_i \in \{+ 1, -1\}$.
However, the variances obtained by SusPs are generally not equal to one~\cite{Yasuda&Tanaka2007}. 
We refer this inconsistency as \textit{the diagonal inconsistency} in this paper.

Herein, we propose a new SusP that overcomes the diagonal inconsistency problem by using \textit{the diagonal trick method}. 
The diagonal trick method is a heuristic technique that originates in the field of the inverse Ising problem~\cite{KR1998, T.Tanaka1998, Yasuda&Tanaka2009}. 
The proposed approach is reduced to the conventional SusP on tree system and 
to the adaptive Thouless-Anderson-Palmer (TAP) equation \cite{Opper&Winther2001a, Opper&Winther2001b} at high temperatures. 

\section{Scheme of Proposed Method} \label{Sec:OurScheme}

\subsection{Gibbs Free Energy and Susceptibility Propagation} \label{SubSec:GFE&SusP}

Let us suppose that an undirected graph $\mcal{G}(V,E)$ with $n$ nodes, where $V = \{1,2,\ldots, n\}$ is the set of all labels of nodes and that $E=\{(i,j)\}$ is the set of all undirected links in the given graph.   
Let us consider an Ising system $\bm{S}=\{S_i \in \{-1,+1\}\mid i = 1,2,\ldots, n\}$, whose Hamiltonian is described as
\begin{align}
\mcal{H}(\bm{S}\mid \bm{h}, \bm{J}):=-\sum_{i=1}^n h_i S_i - \sum_{(i,j) \in E}J_{ij}S_i S_j
\label{eq:Hamiltonian}
\end{align}
on the graph $\mcal{G}(V,E)$, 
where the second summation is the summation taken over all the nearest pairs. 
The parameters $\{J_{ij}\}$, which express the weights of interactions, are symmetrical with respect to $i$ and $j$.
The Gibbs free energy (GFE) of this system, obtained by constraints for local magnetizations, is expressed as~\cite{Yasuda&Tanaka2009}
\begin{align}
G(\bm{m})&:= -  \sum_{i=1}^n h_i m_i + \max_{\bm{\phi}}\Big\{\sum_{i=1}^n \phi_i m_i \nn
\aldef
- \beta^{-1}\ln \sum_{\bm{S}} \exp \Big( - \beta \mcal{H}(\bm{S}\mid \bm{\phi}, \bm{J})\Big)\Big\},
\label{eq:Gibbs}
\end{align} 
where the term $\beta$ denotes the inverse temperature. 
It is known that the minimum point of this GFE with respect to $\bm{m}$ provides the Helmholtz free energy 
of this system, $ F:= - \beta^{-1}\ln \sum_{\bm{S}}\exp  \big(-\beta \mcal{H}(\bm{S}\mid \bm{h}, \bm{J})\big)$, and that the values of 
$\bm{m}=\{m_i \mid i = 1,2,\ldots,n\}$ coincide with the exact local magnetizations of this system at the minimum point, i.e., 
$F = \min_{\bm{m}}G(\bm{m})$ and $\ave{\bm{S}} = \argmin_{\bm{m}}G(\bm{m})$, where the notation $\ave{\cdots}$ describes the (true) expectation in the Ising system (\ref{eq:Hamiltonian}).
Susceptibilities of the Ising system, $\chi_{ij}:= \ave{S_iS_j} - \ave{S_i}\ave{S_j}$, are obtained as 
$ \beta \chi_{ij} = \partial m_i^* /\partial h_j$, where $\bm{m}^*:= \argmin_{\bm{m}} G(\bm{m})$.

SusP is an approximation technique that can be used to evaluate the susceptibilities of a given system
~\cite{K.Tanaka2003, Welling&Teh2003, Welling&Teh2004, Mezard&Mora2009}. 
In the SusP, we employ an approximate GFE, $G_{\mrm{app}}(\bm{m})$, for example, the naive mean-field GFE and the Bethe GFE, instead of the true GFE (\ref{eq:Gibbs}).  
In the SusP technique, the approximate susceptibilities are obtained by $\beta \chi_{ij} \approx \beta \hat{\chi}_{ij}:=\partial \hat{m}_i /\partial h_j$, 
where $\hat{\bm{m}}:= \argmin_{\bm{m}} G_{\mrm{app}}(\bm{m})$. 
SusPs have been also used to address the inverse Ising problem~\cite{Yasuda&Tanaka2009, Marinari&Kerrebroeck2010}. 

\subsection{Diagonal Inconsistency Problem and Proposed Scheme} \label{SubSec:DIP&OurScheme}

Since $S_i \in \{+1,-1\}$, the quantities $\chi_{ii} + (m_i^*)^2 = \ave{S_i^2}$ are obviously equal to one. 
However, it has been known that the quantities obtained by a SusP, i.e., $\hat{\chi}_{ii} + \hat{m}_i^2$, are generally not equal to one due to approximation~\cite{Yasuda&Tanaka2007}. 
In particular, such a diagonal inconsistency can become a serious problem in attempting to address the inverse Ising problem~\cite{Yasuda&Tanaka2009}. 
In order to avoid this diagonal inconsistency, a heuristic technique called the diagonal trick method has been employed 
in various mean-field algorithms for address the inverse Ising problem~\cite{KR1998, T.Tanaka1998, Yasuda&Tanaka2009}. 

In this paper, we propose a new scheme of SusP using the concept of the diagonal trick method in which an obtained SusP satisfies the diagonal consistency. 
The proposed scheme is quite simple and can be described as follows:
first, with controllable parameters $\bm{\Lambda}$, we extend an approximated GFE, $G_{\mrm{app}}(\bm{m})$, as 
\begin{align}
G_{\mrm{app}}^{\dagger}(\bm{m}, \bm{\Lambda}):= G_{\mrm{app}}(\bm{m}) - \frac{1}{2}\sum_{i=1}^n \Lambda_i\big(1- m_i^2\big).
\label{eq:ext-GFEapp}
\end{align}
The second term appears due to the use of the diagonal trick method~\cite{Yasuda&Tanaka2009}. 
Next, we minimize the extended approximate GFE in Eq. (\ref{eq:ext-GFEapp}) with respect to $\bm{m}$ and formulate a SusP in the general manner as 
\begin{align}
\beta \chi_{ij}^{\dagger}(\bm{\Lambda}):= \partial m_i^{\dagger}(\bm{\Lambda}) /\partial h_j, 
\label{eq:ext-SusP}
\end{align}
where $\bm{m}^{\dagger}(\bm{\Lambda}):= \argmin_{\bm{m}} G_{\mrm{app}}^{\dagger}(\bm{m}, \bm{\Lambda})$.
The values of $\bm{m}^{\dagger}(\bm{\Lambda})$ and $\bm{\chi}_{ij}^{\dagger}(\bm{\Lambda})$ depend on the values of $\bm{\Lambda}$. 
Finally, we determine the values of $\bm{\Lambda}$ so as to satisfy the diagonal consistencies by using
\begin{align}
\chi_{ii}^{\dagger}(\bm{\Lambda}) + m_i^{\dagger}(\bm{\Lambda})^2 = 1.
\label{eq:DiagConsist}
\end{align}
This equation is termed \textit{the diagonal matching equation} in this paper, 
because it leads to the matching of the diagonals of the susceptibilities (plus the square of local magnetizations) 
obtained by an approximation with those obtained by an exact evaluation. 
An alternative interpretation of this equation is provided in Appendix \ref{app:review-DME}.

\section{Improved Susceptibility Propagation with Bethe Approximation} \label{Sec:A-SusP}

In the context of SusP, the Bethe GFE is generally used as the approximate GFE, i.e. we use $G_{\mrm{app}}(\bm{m})$~\cite{Mezard&Mora2009, Marinari&Kerrebroeck2010}.
Herein, we explicitly formulate a SusP obtained from our proposed scheme by employing the Bethe GFE.

\subsection{Belief Propagation with $\bm{\Lambda}$} \label{SubSec:BP}

In the Ising system, the Bethe GFE is expressed as~\cite{Horiguchi1981, Welling&Teh2003, Yasuda&Tanaka2009}
\begin{align}
&G_{\mrm{app}}(\bm{m})=-  \sum_{i=1}^n h_i m_i -\sum_{(i,j) \in E}J_{ij} \xi_{ij} \nn
&+ \frac{1}{\beta}\sum_{i=1}^n (1- |\partial i|) \sum_{S_i = \pm 1}\rho_1(S_i\mid m_i)\ln \rho_1(S_i\mid m_i)\nn
&+ \frac{1}{\beta}\sum_{(i,j) \in E} \sum_{S_i,S_j = \pm 1} \rho_2(S_i, S_j\mid m_i, m_j, \xi_{ij})\nn
&\times \ln \rho_2(S_i, S_j\mid m_i, m_j, \xi_{ij}),
\label{eq:BetheGFE}
\end{align}
where
\begin{align*}
&\rho_1(S_i\mid m_i):= \frac{1 + S_i m_i}{2}, \nn
&\rho_2(S_i, S_j\mid m_i, m_j, \xi_{ij}):= \frac{1 + S_i m_i + S_j m_j + S_iS_j \xi_{ij}}{4},
\end{align*}
and 
\begin{align}
\xi_{ij}&:= \coth(2 \beta J_{ij})\Big\{1-\Big(1-(1-m_i^2-m_j^2) \tanh^{2}(2 \beta J_{ij}) \nn
\aldef - 2m_im_j\tanh(2 \beta J_{ij})\Big)^{1/2} \Big\}. 
\label{eq:BP-cov}
\end{align}
The notation $\partial i$ denotes the set of nodes connecting to node $i$, i.e., $\partial i:= \{ j \mid (i,j) \in E\}$. 

In accordance with Eq. (\ref{eq:ext-GFEapp}), we extend the Bethe GFE by adding the diagonal trick term, and we minimize the extended GFE, 
$G_{\mrm{app}}^{\dagger}(\bm{m}, \bm{\Lambda})$, with respect to $\bm{m}$. 
By using a similar manipulation in references \cite{Horiguchi1981, Yasuda&Tanaka2009}, for given a set of parameters $\bm{\Lambda}$, 
we can obtain $\bm{m}^{\dagger}(\bm{\Lambda})$ as the solutions to a set of simultaneous equations given by
\begin{align}
\mcal{M}_{i\rightarrow j}(\bm{\Lambda})&=\tanh^{-1}\Big\{ \tanh( \beta J_{ij}) 
\tanh\Big( \beta h_i \nn
\aleq- \beta \Lambda_i m_i^{\dagger}(\bm{\Lambda}) + \sum_{k \in \partial i \setminus \{j\}} \mcal{M}_{k\rightarrow i}(\bm{\Lambda})\Big)\Big\}
\label{eq:message-passing}
\end{align}
and
\begin{align}
m_i^{\dagger}(\bm{\Lambda})=\tanh\Big(\beta h_i- \beta \Lambda_i m_i^{\dagger}(\bm{\Lambda}) + \sum_{j \in \partial i } \mcal{M}_{j\rightarrow i}(\bm{\Lambda})\Big).
\label{eq:det-m_dagger}
\end{align}
Equation (\ref{eq:message-passing}) is known as the message-passing rule of belief propagation. 
The quantity $\mcal{M}_{i\rightarrow j}(\bm{\Lambda})$ is interpreted as the message, or the effective field, propagated from node $i$ to node $j$ through the link $(i,j)$. 
If $\bm{\Lambda}=\bm{0}$, Eqs. (\ref{eq:message-passing}) and (\ref{eq:det-m_dagger}) are reduced to the usual belief propagation for Ising systems.

\subsection{SusP with $\bm{\Lambda}$ and Diagonal Matching}

By using Eq. (\ref{eq:ext-SusP}) and $\bm{m}^{\dagger}(\bm{\Lambda})$ obtained from Eqs. (\ref{eq:message-passing}) and (\ref{eq:det-m_dagger}), 
we can formulate a SusP for a given parameter set $\bm{\Lambda}$ as follows:
\begin{align}
\beta \chi_{ij}^{\dagger}(\bm{\Lambda})=\frac{1 - m_i^{\dagger}(\bm{\Lambda})^2}{1 + \beta \Lambda_i \big(1 - m_i^{\dagger}(\bm{\Lambda})^2\big)}
\Big(\beta \delta_{i,j} + \sum_{k \in \partial i}\eta_{k\rightarrow i}^{(j)}(\bm{\Lambda})\Big),
\label{eq:ad-SusP}
\end{align}
where $\delta_{i,j}$ is the Kronecker delta. 
The quantities $\eta_{i\rightarrow j}^{(k)} (\bm{\Lambda})$ are defined as 
$\eta_{i\rightarrow j}^{(k)} (\bm{\Lambda}) := \partial \mcal{M}_{i\rightarrow j}(\bm{\Lambda})/\partial h_k$, 
and they satisfy the relation
\begin{widetext}
\begin{align}
\eta_{i\rightarrow j}^{(k)} (\bm{\Lambda})
=\frac{
\sinh(2\beta J_{ij})\Big(\beta \delta_{i,k} - \beta^2 \Lambda_i \chi_{ik}^{\dagger}(\bm{\Lambda}) 
+ \sum_{l \in \partial i \setminus \{j\}}\eta_{l\rightarrow i}^{(k)}(\bm{\Lambda}) \Big)
}
{
\cosh (2\beta J_{ij})  
+ \cosh 2\Big(\beta h_i - \beta \Lambda_i m_i^{\dagger}(\bm{\Lambda})
+ \sum_{l \in \partial i \setminus \{j\}}\mcal{M}_{l\rightarrow i}(\bm{\Lambda}) \Big)
}.
\label{eq:ad-SusP_sub}
\end{align}
\end{widetext}
Eqs. (\ref{eq:ad-SusP}) and (\ref{eq:ad-SusP_sub}) are obtained by differentiating Eq. (\ref{eq:det-m_dagger}) with respect to $h_j$ and Eq. (\ref{eq:message-passing}) with respect to $h_k$, 
respectively.
If $\bm{\Lambda}=\bm{0}$, Eqs. (\ref{eq:ad-SusP}) and (\ref{eq:ad-SusP_sub}) are reduced to the usual SusP for Ising systems. 

To determine the values of $\bm{\Lambda}$ that satisfy the diagonal consistencies, we substitute Eq. (\ref{eq:DiagConsist}) in Eq. (\ref{eq:ad-SusP}), which leads to the relations
\begin{align}
\Lambda_i = \frac{1}{\beta^2}\frac{1}{1-m_i^{\dagger}(\bm{\Lambda})^2} \sum_{j \in \partial i}\eta_{j \rightarrow i}^{(i)}(\bm{\Lambda}).
\label{eq:DiagConsis_ad-SusP}
\end{align}

The proposed SusP, referred to as \textit{the improved susceptibility propagation} (I-SusP) in this paper, 
can be computed by simultaneously solving Eqs. (\ref{eq:message-passing}), (\ref{eq:det-m_dagger}), 
(\ref{eq:ad-SusP}), (\ref{eq:ad-SusP_sub}), and (\ref{eq:DiagConsis_ad-SusP}).   

From the arguments given in the last paragraph in Appendix \ref{app:review-DME}, on a tree graph, 
the proposed I-SusP is equivalent to the ordinary SusP and gives an exact solution, 
because the Bethe GFE coincides with the true GFE on a tree graph.  

\subsection{Connection to Adaptive TAP Equation}

The proposed I-SusP given in the previous section can be viewed as an extension of the adaptive TAP equation~\cite{Opper&Winther2001a, Opper&Winther2001b}. 
The Bethe GFE in Eq. (\ref{eq:BetheGFE}) can be expanded as~\cite{Horiguchi1981}
\begin{align}
\beta G_{\mrm{app}}(\bm{m})&=-  \beta \sum_{i=1}^n h_i m_i - \beta \sum_{(i,j) \in E}J_{ij} m_im_j \nn
&+ \sum_{i=1}^n  \sum_{S_i = \pm 1}\rho_1(S_i\mid m_i)\ln \rho_1(S_i\mid m_i) + O(\beta^2).
\label{eq:MF-GFE}
\end{align}
This expression is known as the naive mean-field GFE.

By using the naive mean-field GFE instead of the Bethe GFE in our scheme, we can derive the following simultaneous equations:
\begin{align}
m_i^{\dagger}(\bm{\Lambda})&=\tanh \beta \Big( h_i + \sum_{j \in \partial i}J_{ij}m_j^{\dagger}(\bm{\Lambda}) - \Lambda_i m_i^{\dagger}(\bm{\Lambda}) \Big), 
\label{eq:MFE}\\
\beta \chi_{ij}^{\dagger}(\bm{\Lambda})&=\frac{1 - m_i^{\dagger}(\bm{\Lambda})^2}{1 + \beta \Lambda_i \big(1 - m_i^{\dagger}(\bm{\Lambda})^2\big)}\nn
\aleq\times \Big(\beta \delta_{i,j} + \beta^2 \sum_{k \in \partial i}J_{ik} \chi_{kj}^{\dagger}(\bm{\Lambda}) \Big),
\label{eq:MFE-SusP}
\end{align}
and 
\begin{align}
\Lambda_i = \frac{1}{1-m_i^{\dagger}(\bm{\Lambda})^2} \sum_{k \in \partial i}J_{ik} \chi_{ki}^{\dagger}(\bm{\Lambda}).
\label{eq:DiagConsis-MF}
\end{align}
Eq. (\ref{eq:MFE}), which is known as the mean-field equation, is obtained from the extremal condition of Eq. (\ref{eq:MF-GFE}) with respect to $\bm{m}$. 
Eqs. (\ref{eq:MFE-SusP}) and (\ref{eq:DiagConsis-MF}) are obtained by using Eqs. (\ref{eq:ext-SusP}) and (\ref{eq:DiagConsist}), respectively.

Alternative representations of Eqs. (\ref{eq:MFE-SusP}) and (\ref{eq:DiagConsis-MF}) are possible. 
From Eq. (\ref{eq:LRR-app}), we find that the mean-field susceptibility $\beta \chi_{ij}^{\dagger}(\bm{\Lambda})$ in Eq. (\ref{eq:MFE-SusP})
is equivalent to the $ij$-th element of the Hessian matrix of the extended mean-field GFE that is obtained by adding the diagonal trick term to Eq. (\ref{eq:MF-GFE}) 
in accordance with to Eq. (\ref{eq:ext-GFEapp}):
\begin{align}
\beta \chi_{ij}^{\dagger}(\bm{\Lambda}) = \big( \bm{D} - \bm{J} \big)^{-1}_{ij}, 
\label{eq:MF-LRR}
\end{align}
where the term $\bm{J}$ denotes the matrix whose $ij$-th element is $J_{ij}$ and the term $\bm{D}$ denotes the diagonal matrix whose $ii$-th element is defined by 
$(\bm{D})_{ii}:= \beta^{-1}(1 - m_i^{\dagger}(\bm{\Lambda})^2)^{-1} + \Lambda_i$. 
Eq. (\ref{eq:MF-LRR}) can be also directly obtained from Eq. (\ref{eq:MFE-SusP}).
From Eq. (\ref{eq:DiagConsist}) and (\ref{eq:MF-LRR}), we obtain 
\begin{align}
\big( \beta \bm{D} - \beta \bm{J} \big)^{-1}_{ii} = 1 - m_i^{\dagger}(\bm{\Lambda})^2. 
\label{eq:DiagConsis-MF-2}
\end{align}
Eqs. (\ref{eq:MF-LRR}) and (\ref{eq:DiagConsis-MF-2}) are the alternative representations of Eqs. (\ref{eq:MFE-SusP}) and (\ref{eq:DiagConsis-MF}), respectively. 

By using the shift $\beta \Lambda_i  + (1 - m_i^{\dagger}(\bm{\Lambda})^2)^{-1} \rightarrow \beta \Lambda_i$, 
Eqs. (\ref{eq:MFE}) and (\ref{eq:DiagConsis-MF-2}) become
\begin{align}
m_i^{\dagger}(\bm{\Lambda})&=\tanh \beta \bigg( h_i + \sum_{j \in \partial i}J_{ij}m_j^{\dagger}(\bm{\Lambda}) - \Lambda_i m_i^{\dagger}(\bm{\Lambda})  \nn
\aleq
+ \frac{m_i^{\dagger}(\bm{\Lambda})}{\beta( 1 - m_i^{\dagger}(\bm{\Lambda})^2 )}\bigg)
\label{eq:ADTAPE-1}
\end{align}
and 
\begin{align}
\big( \beta \bm{D}_{\bm{\Lambda}} - \beta \bm{J} \big)^{-1}_{ii} = 1 - m_i^{\dagger}(\bm{\Lambda})^2, 
\label{eq:ADTAPE-2}
\end{align}
respectively, where the term $\bm{D}_{\bm{\Lambda}}$ denotes the diagonal matrix whose $ii$-th element is $\Lambda_i$.
Equations (\ref{eq:ADTAPE-1}) and (\ref{eq:ADTAPE-2}) coincide with the adaptive TAP equation. 
Therefore, we find that our I-SusP is equivalent to the adaptive TAP equation when $\beta$ is extremely small.

Based on the arguments presented in this section, we can reinterpret the adaptive TAP equation as a SusP based on the naive mean-field approximation with the diagonal trick method. 
In the context of the adaptive TAP equation, the parameters $\bm{\Lambda}$ are regarded as effects of \textit{the Onsager reaction term}. 
Hence, we can expect that our diagonal trick term behaves as an Onsager reaction term.

\subsection{Discussion}

\begin{figure}[hbt]
\begin{center}
\includegraphics[height=4.0cm]{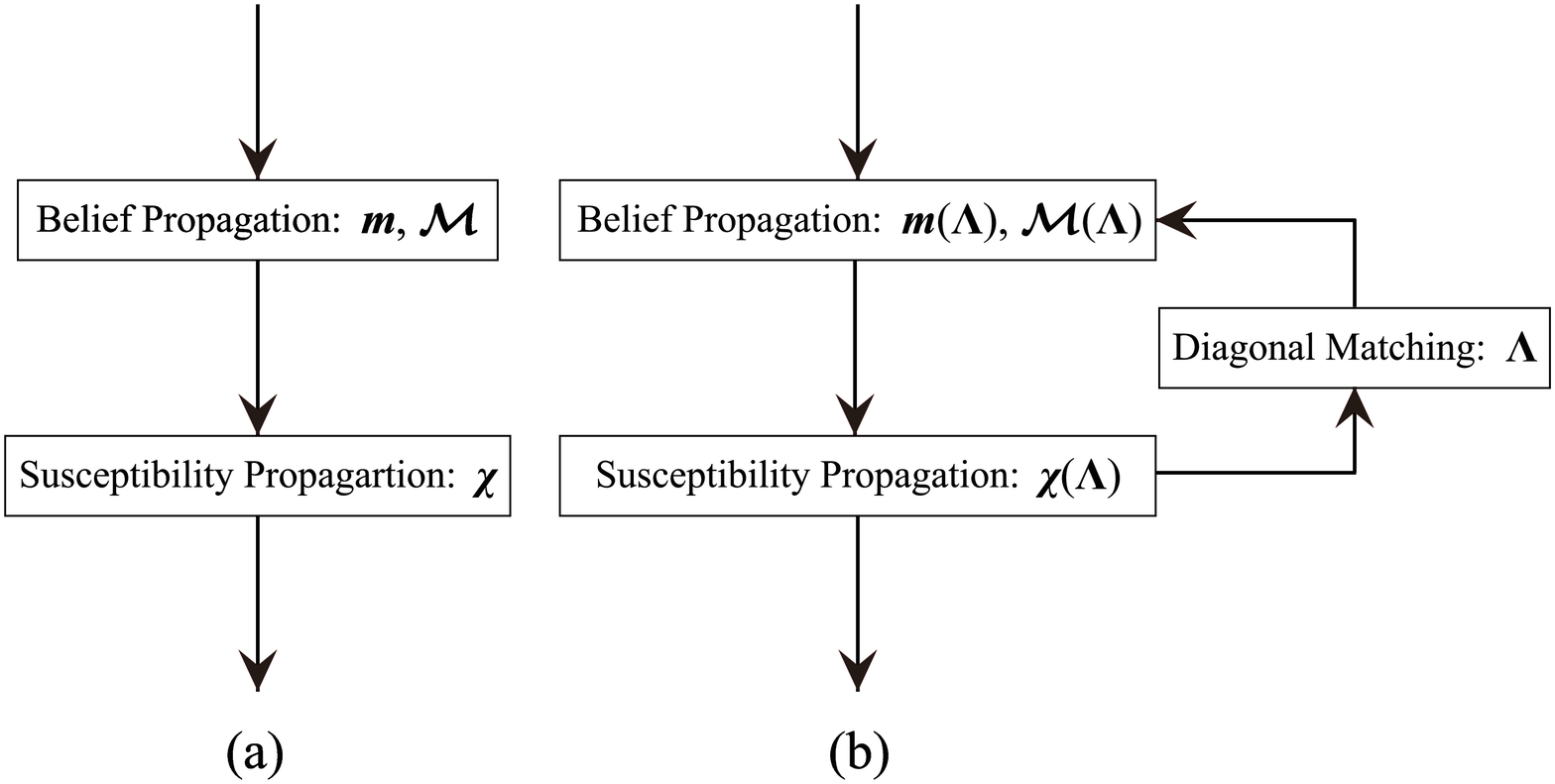}
\end{center}
\caption{Illustration of schemes of (a) conventional SusP and (b) I-SusP.}
\label{fig:Scheme}
\end{figure}

In the conventional SusP scheme, for a given system, we first solve a belief propagation (Eqs. (\ref{eq:message-passing}) and (\ref{eq:det-m_dagger}) with $\bm{\Lambda} = \bm{0}$), 
and we obtain solutions to the belief propagation. These solutions are the values of the local magnetizations and the messages.  
Subsequently, to obtain the susceptibilities, we solve a SusP (Eqs. (\ref{eq:ad-SusP}) and (\ref{eq:ad-SusP_sub}) with $\bm{\Lambda} = \bm{0}$) by using the solutions of the belief propagation. 
Therefore, in the conventional scheme shown to the left in Fig. \ref{fig:Scheme}, the results of the SusP cannot affect the belief propagation.
In contrast, the proposed scheme shown to the right in Fig. \ref{fig:Scheme} includes feedback from the SusP to the belief propagation through $\bm{\Lambda}$. 
This feedback marks a significant difference between the conventional scheme and the proposed scheme. 
If one employs a synchronous updating algorithm, the computational cost of solving the usual SusP is $O(n |E|)$. 
Since Eq. (\ref{eq:DiagConsis_ad-SusP}) does not contribute to the order of the computational cost, when using a synchronous updating algorithm, 
the computational cost of solving the I-SusP is the same as that of solving the usual SusP.

The Bethe approximation, which is equivalent to the belief propagation, 
is regarded as the approximation that neglects the effects of all the loops in a given system~\cite{Yasuda&Tanaka2006}. 
It is known that, by applying a SusP to the Bethe approximation in the conventional manner, the effects of the loops neglected in the Bethe approximation are partially restored 
and that the restored loop effects improve the accuracy of the approximation of susceptibilities~\cite{Yasuda&Tanaka2007}. 
Hence, it can be expected that the feedback in our scheme restores the effects of the loops, 
revived by a SusP, to a belief propagation, and that this improves not only the SusP but also the belief propagation. 
Indeed, a similar observation has been made in reference \cite{T.Tanaka1998} and the validity of the observation was examined in the study from a perturbative point of view.

\section{Numerical Experiment} \label{Sec:NumExp}

In this section, we provide the results of certain numerical experiments. 
To measure the performances of our approximations, we use the mean absolute distances (MADs) defined by
\begin{align}
D_{V}&:= \frac{1}{n} \sum_{i =1}^n| \ave{S_i} - \ave{S_i}_{\mrm{app}}|, \\
D_{E}&:= \frac{1}{|E|} \sum_{(i,j) \in E}| \ave{S_iS_j} - \ave{S_iS_j}_{\mrm{app}}|,
\end{align}
and 
\begin{align}
D_{E^*}&:= \frac{1}{n(n-1)/2 - |E|} \sum_{(i,j) \not \in E}| \ave{S_iS_j} - \ave{S_iS_j}_{\mrm{app}}|,
\end{align}
where the notation $\ave{\cdots}_{\mrm{app}}$ indicates the average value evaluated by an approximation method.
The expression $D_{E}$ denotes the MAD for correlation functions between nearest-neighbor pairs 
and the expression $D_{E^*}$ denotes the MAD for correlation functions between non-nearest-neighbor pairs. 
In the following experiments, the parameters $\bm{h}$ and $\bm{J}$ are independently obtained from the distributions $p_h(h_i)$ and $p_J(J_{ij})$
and the inverse temperature $\beta$ is set to the value of one.

Let us consider an Ising system on a graph $\mcal{G}(V,E)$, with $n = 22$ and 
\begin{align}
p_h(h_i)&=\mcal{N}(h_i \mid 0.1^2), \\
p_J(J_{ij})&= (1-p)\delta( J_{ij}) + p\mcal{N}(J_{ij} \mid \sigma^2 / (pn) ),
\end{align}
where $\delta(x)$ denotes the Dirac delta function and the expression $\mcal{N}(x \mid \sigma^2)$ represents the standard Gaussian with variance $\sigma^2$. 
In this case, this graph is regarded as a random graph (known as the Erd{\H o}s-R{\'e}nyi graph) with the connection probability $p$ 
(probability of the link $(i,j)$ being presented between nodes $i$ and $j$). 
This system is identified as the Sherrington-Kirkpatrick (SK) model when $p=1$.

\begin{figure*}[hbt]
\begin{center}
\includegraphics[height=4.0cm]{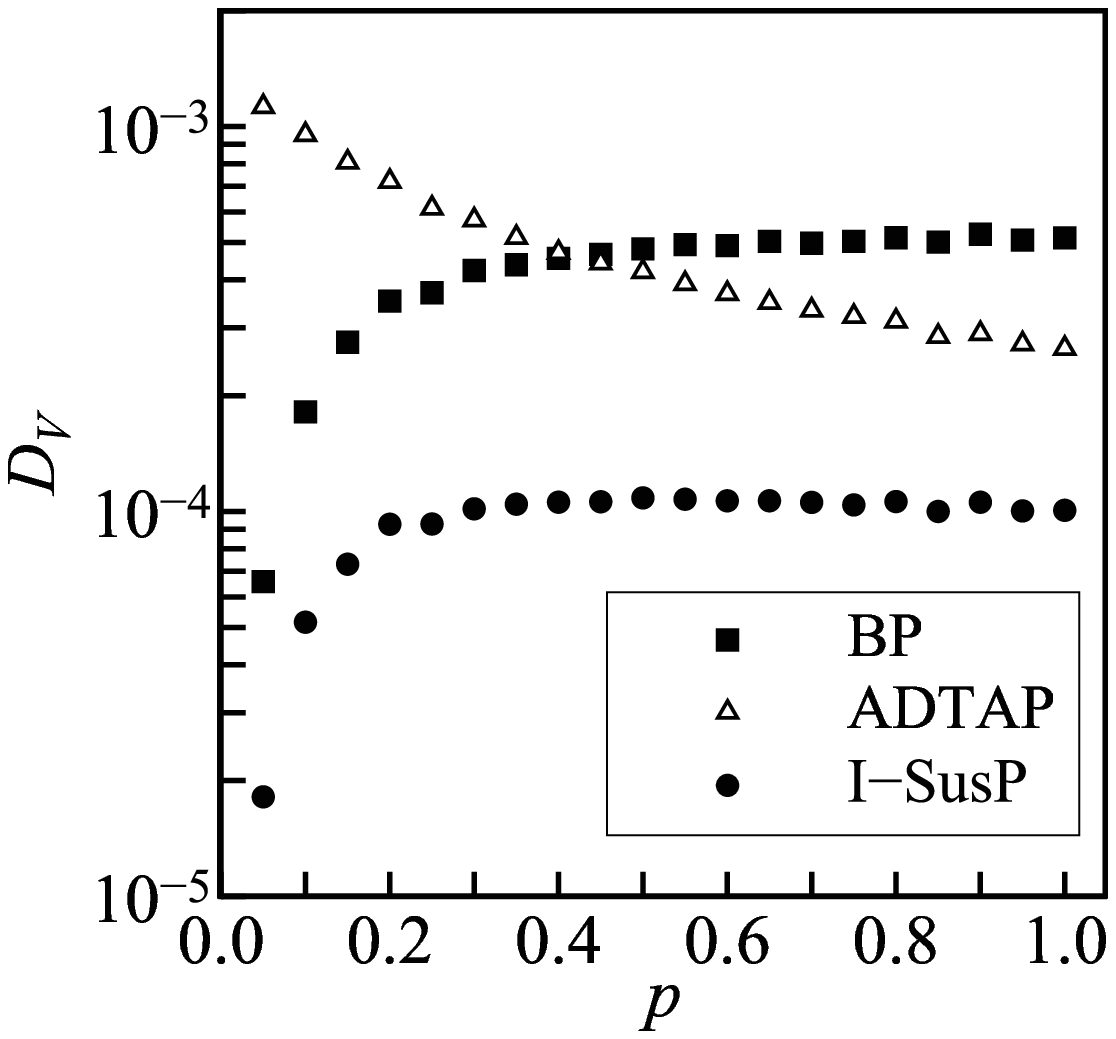}
\includegraphics[height=4.0cm]{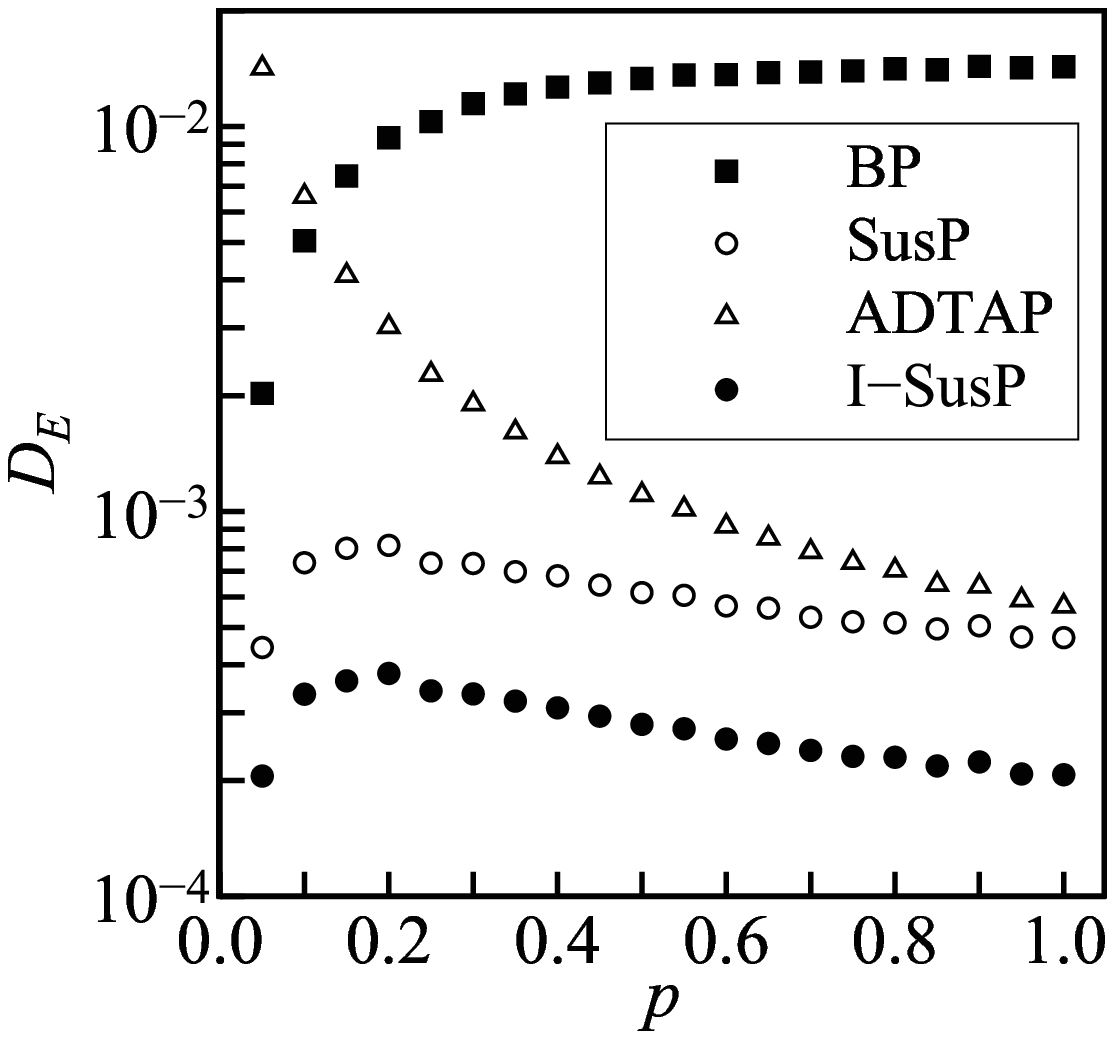}
\includegraphics[height=4.0cm]{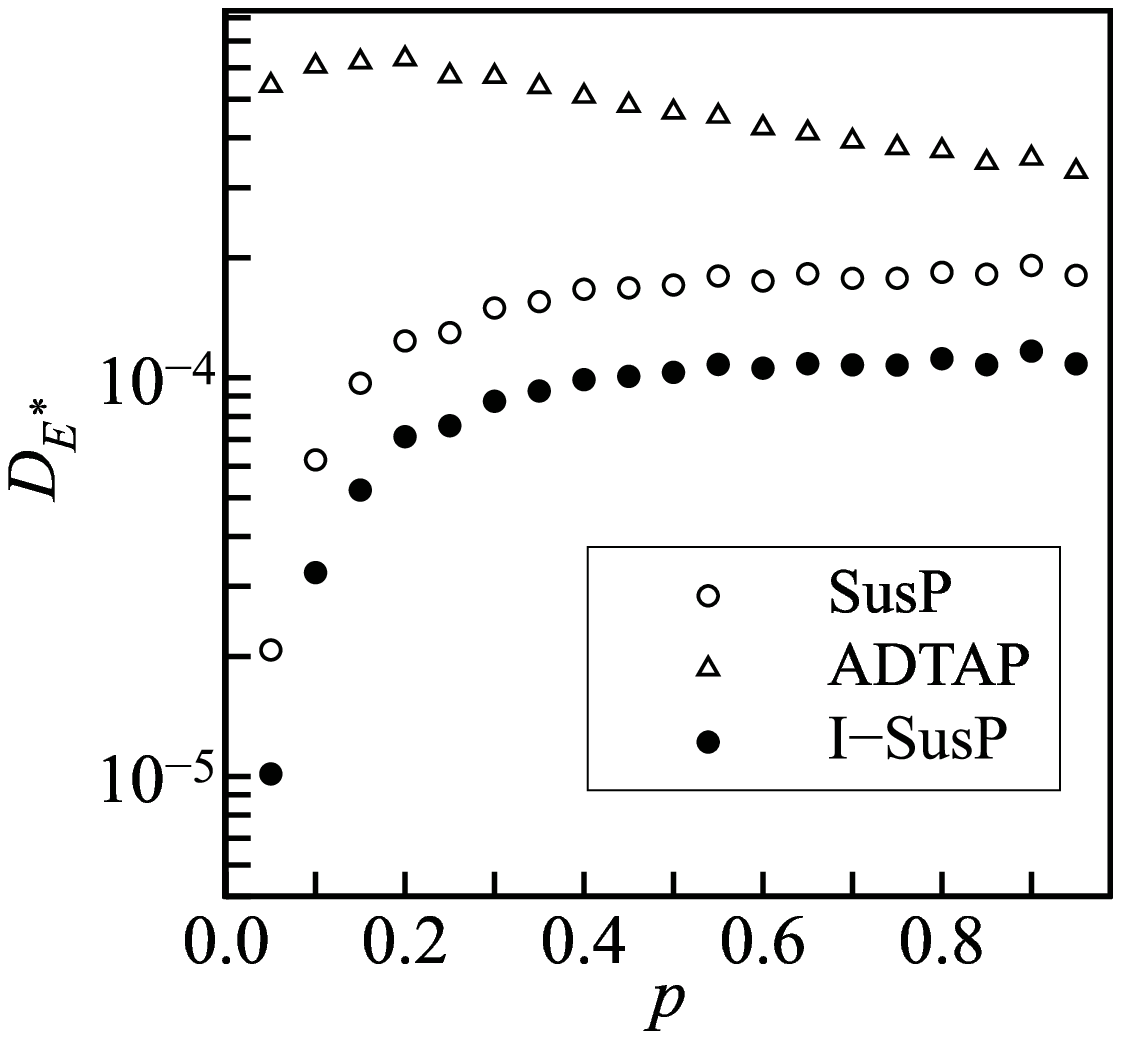}
\end{center}
\caption{MADs when $\sigma = 0.3$. Each point is obtained by averaging over 1500 trials.}
\label{fig:ER0.3}
\end{figure*}

\begin{figure*}[hbt]
\begin{center}
\includegraphics[height=4.0cm]{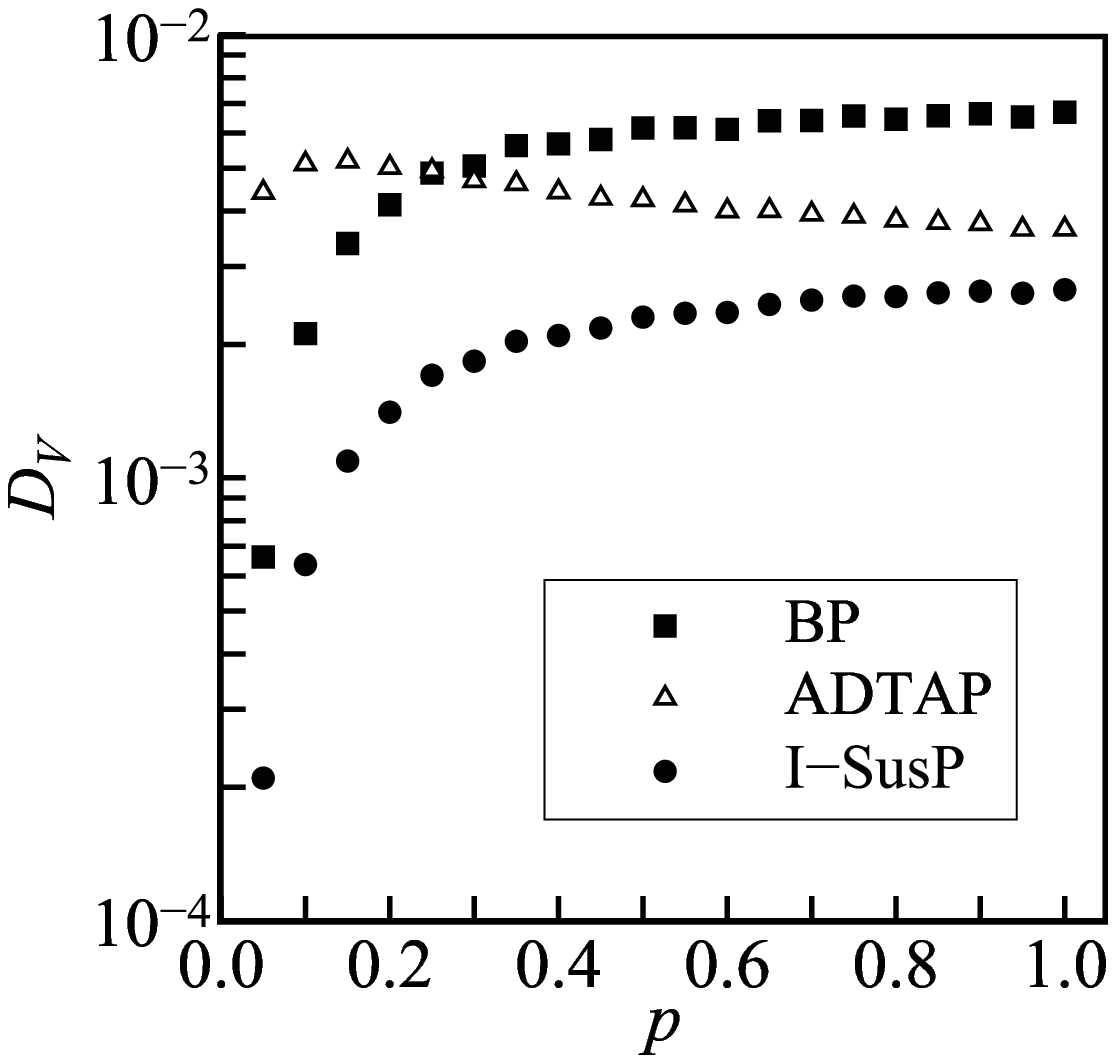}
\includegraphics[height=4.0cm]{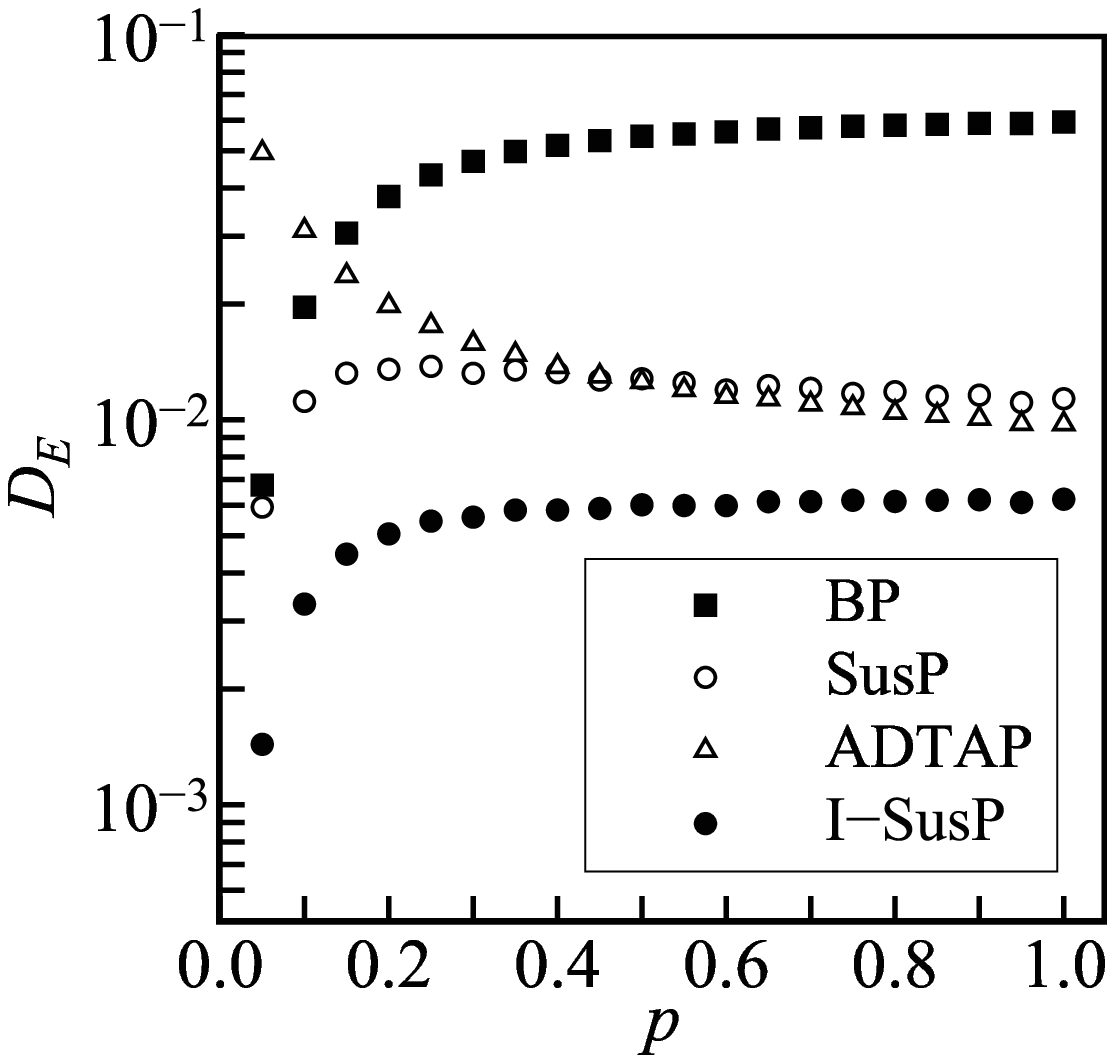}
\includegraphics[height=4.0cm]{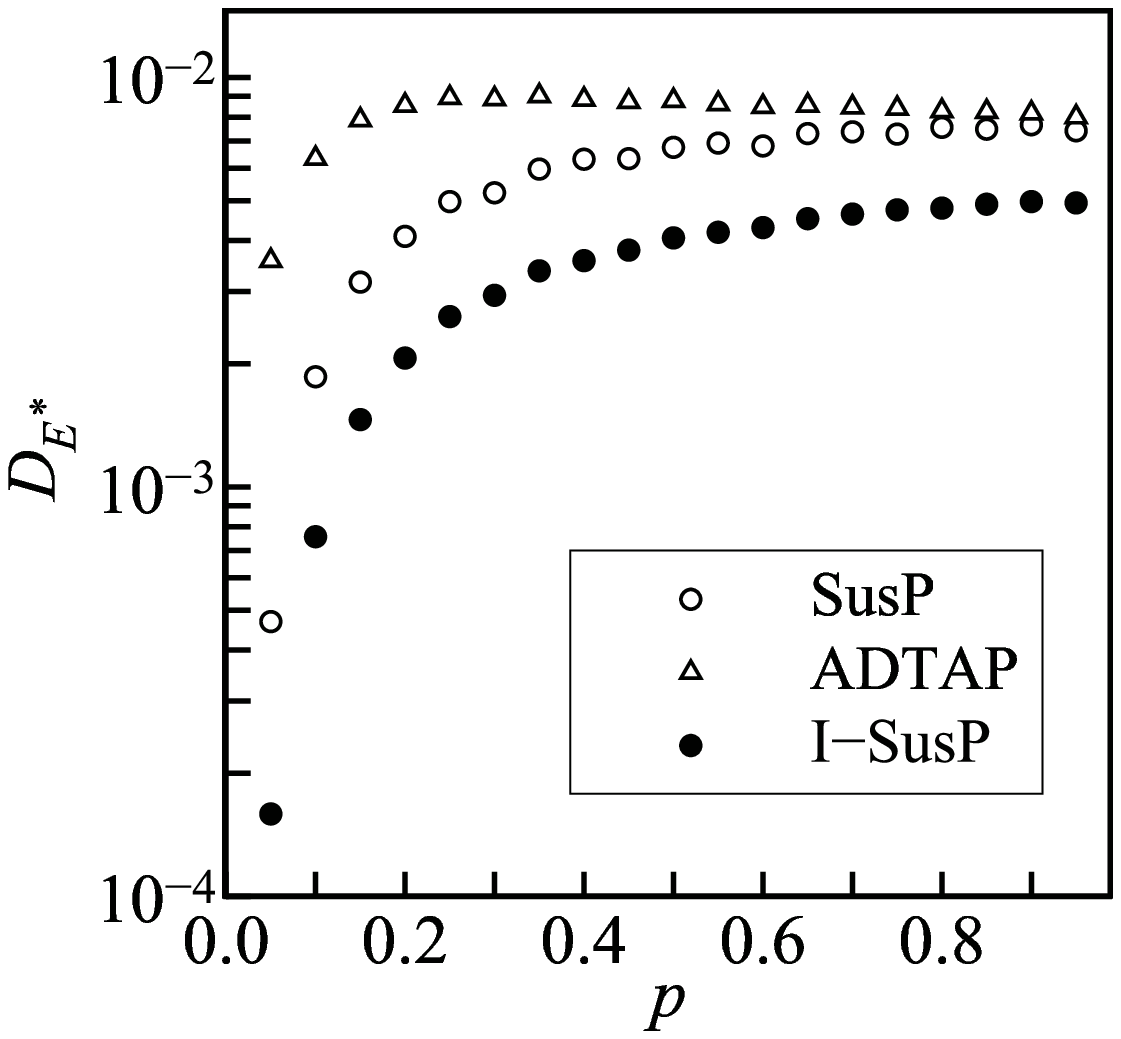}
\end{center}
\caption{MADs when $\sigma = 0.6$. Each point is obtained by averaging over 10000 trials.}
\label{fig:ER0.6}
\end{figure*}

Figs. \ref{fig:ER0.3} and \ref{fig:ER0.6} show the plots of MADs ($D_V$, $D_E$, and $D_{E^*}$) 
against $p$ when $\sigma = 0.3$ and $\sigma = 0.6$, respectively.
`The curves indicated by `BP'', ``SusP,'' and ``ADTAP'' in these plots are the results obtained via the normal belief propagation, the normal SusP, 
and the adaptive TAP equation, respectively. 
It is to be noted that in the framework of the Bethe approximation (belief propagation), 
the correlations $\{\ave{S_iS_j}\}$ between nearest-neighbors can be approximated by Eq. (\ref{eq:BP-cov}): 
$\ave{S_iS_j}\approx \xi_{ij}$~\cite{Horiguchi1981}. 
The curves indicated by``I-SusP'' in these plots are the results obtained by the proposed I-SusP.

In the region where $p$ is small, the belief propagation shows good performance because it is an appropriate for sparse graphs. 
On the other hand, the adaptive TAP equation shows good performance in the region where $p$ is large,  
because the adaptive TAP equation is based on the mean-field method for mean-field models of dense graphs such as SK models. 
Our method shows the best performance over all value of $p$.
As mentioned previously, our method includes the normal SusP and the adaptive TAP as special cases. 
Hence, our method can be a good approximation for both: sparse and dense systems, and it exhibits a robust performance over a wide range of values of $p$.

\section{Conclusion}

In this paper, we proposed a new SusP scheme by incorporating the concept of the diagonal trick method in the conventional SusP scheme. 
The I-SusP obtained from the proposed scheme includes the conventional SusP and the adaptive TAP equation as special cases, 
and it exhibits a robust performance for various types of network structures of given systems, 
as observed from the results of our numerical experiments presented in Sec. \ref{Sec:NumExp}. 
In principle, with the I-SusP, different types of Hamiltonians can be obtained using Eq. (\ref{eq:Hamiltonian})
, e.g., those representing Ising systems with higher-order interactions, and 
different types of approximations from the belief propagation, e.g., generalized belief propagations, can be addressed using our proposed scheme. 

In our scheme, it is necessary that all random variables assume binary values of $\pm 1$. 
Hence, the present scheme cannot immediately be applied to systems with multivalued variables such as $Q$-Ising systems. 
Therefore, it is important to extend the present scheme to multivalued systems.
This topic will be considered in future studies.

\subsection*{Acknowledgment}
This work was partly supported by Grants-In-Aid (No.21700247, No.24700220, and No. 22300078) 
for Scientific Research from the Ministry of Education, Culture, Sports, Science and Technology, Japan.
The authors thank Dr. Cyril Furtlehner for discussions at various stages of this work.

\appendix

\section{Linear Response Relation for Approximate GFEs}

Let us consider a function expressed in the form
\begin{align*}
G_0(\bm{m}) = -  \sum_{i=1}^n h_i m_i + g_0(\bm{m}).
\end{align*}
It is be to note that this expression includes all the GFEs (the true GFE, the approximate GFEs, and the extended approximate GFEs) presented in this paper.
At the minima of the function, from its extremal condition, the relation
\begin{align}
- h_i + \left. \frac{\partial g_0(\bm{m})}{\partial m_i} \right|_{\bm{m} = \bm{m}^{\star}} = 0
\label{eq:extre-condition}
\end{align}
holds, where $\bm{m}^{\star}:= \argmin_{\bm{m}} G_0(\bm{m})$. 
By differentiating Eq. (\ref{eq:extre-condition}) with respect to $h_j$, we obtain
\begin{align*}
\delta_{i,j} &=  \sum_{k=1}^n\left. \frac{\partial^2 g_0(\bm{m})}{\partial m_i \partial m_k} \right|_{\bm{m} = \bm{m}^{\star}} \frac{\partial m_k^{\star}}{\partial h_j} \nn
&=\sum_{k=1}^n\left. \frac{\partial^2 G_0(\bm{m})}{\partial m_i \partial m_k} \right|_{\bm{m} = \bm{m}^{\star}} \frac{\partial m_k^{\star}}{\partial h_j}.
\end{align*} 
This expression indicates that the relation 
\begin{align}
\bm{H}_0^{-1} = \bm{\chi}^{\star}
\label{eq:LRR-app}
\end{align}
holds, where the term $\bm{H}_0$ denotes the Hessian matrix of $G_0(\bm{m})$ whose $ij$-th element is defined by 
$(\bm{H}_0)_{ij} := \partial^2 G_0(\bm{m})/\partial m_i \partial m_j |_{\bm{m} = \bm{m}^{\star}}$ 
and the term $\bm{\chi}^{\star}$ denotes the matrix whose $ij$-th element is defined by
$(\bm{\chi}^{\star})_{ij} := \partial m_i^{\star}/\partial h_j$.

\section{Variational Interpretation of Diagonal Matching Equation} \label{app:review-DME}

The diagonal matching equation in Eq. (\ref{eq:DiagConsist}) is introduced to overcome the diagonal inconsistency problem. 
In this appendix, we provide an alternative interpretation of the equation.

First, we define a measure of closeness between two positive definite symmetric matrices, $\bm{A} \in \mathbb{R}^{n\times n}$ and $\bm{B} \in \mathbb{R}^{n\times n}$, 
in terms of the Kullback-Leibler divergence as
\begin{align}
D(\bm{A}||\bm{B}):=\int_{-\infty}^{\infty}\diff\bm{x}\> \mcal{N}_n(\bm{x} \mid \bm{A})\ln \frac{ \mcal{N}_n(\bm{x} \mid \bm{A})}{\mcal{N}_n(\bm{x} \mid \bm{B})},
\label{eq:KLD}
\end{align}
where 
\begin{align*}
\mcal{N}_n(\bm{x} \mid \bm{\Sigma}):=\frac{1}{\sqrt{(2\pi)^{n} \det(\bm{\Sigma}^{-1})}}\exp\Big(-\frac{1}{2}\bm{x}^{\mrm{t}}\bm{\Sigma}\bm{x}\Big)
\end{align*}
represents the $n$-dimensional standard Gaussian with the covariant matrix $\bm{\Sigma}^{-1}$. 
From the properties of the Kullback-Leibler divergence, $D(\bm{A}||\bm{B}) >0$ and $D(\bm{A}||\bm{B}) = 0$ iff $\bm{A}=\bm{B}$ is ensured.

Let us define the Hessian matrices of the true GFE and $G_{\mrm{app}}^{\dagger}(\bm{m}, \bm{\Lambda})$ as
$\big(\bm{H}(\bm{m}) \big)_{ij}:= \partial^2 G(\bm{m})/ \partial m_i\partial m_j$ 
and as $\big(\bm{H}^{\dagger}(\bm{m},\bm{\Lambda}) \big)_{ij}:= \partial^2 G_{\mrm{app}}^{\dagger}(\bm{m}, \bm{\Lambda})/\partial m_i\partial m_j$,
respectively, and further, let us consider the quantity $D(\bm{H}(\bm{m})||\bm{H}^{\dagger}(\bm{m},\bm{\Lambda}))$ for given values of $\bm{m}$. 
Minimizing $D(\bm{H}(\bm{m})||\bm{H}^{\dagger}(\bm{m},\bm{\Lambda}))$ corresponds to minimizing the (quasi-) distance~\footnote{
This measure is generally not a distance in a precise mathematical sense, because $D(\bm{A}||\bm{B}) \not= D(\bm{B}||\bm{A})$.
} between the true Hessian matrix and the approximate Hessian matrix. 
The minimum condition of $D(\bm{H}(\bm{m})||\bm{H}^{\dagger}(\bm{m},\bm{\Lambda}))$ with respect to $\bm{\Lambda}$ is 
\begin{align}
1 - m_i^2 = \big( \bm{H}^{\dagger}(\bm{m},\bm{\Lambda})^{-1} \big)_{ii},
\label{eq:MinCondition-Lambda}
\end{align}
where we use the fact that $\big( \bm{H}(\bm{m})^{-1} \big)_{ii} = 1 - m_i^2$ holds for any $\bm{m}$. 
When $\bm{m} = \bm{m}^{\dagger}(\bm{\Lambda})$, upon using Eq. (\ref{eq:LRR-app}), Eq. (\ref{eq:MinCondition-Lambda}) yields Eq. (\ref{eq:DiagConsist}). 
Therefore, we can reinterpret the diagonal matching equation as the condition of minimization of distance between the true Hessian matrix and its approximation in terms of the Kullback-Leibler divergence 
at $\bm{m} = \bm{m}^{\dagger}(\bm{\Lambda})$.

When $G_{\mrm{app}}(\bm{m}) = G(\bm{m})$, minimization of $D(\bm{H}(\bm{m})||\bm{H}^{\dagger}(\bm{m},\bm{\Lambda}))$ obviously leads to $\bm{\Lambda}=\bm{0}$. 
This indicates that the diagonal matching equation in Eq. (\ref{eq:DiagConsist}) yields the result $\bm{\Lambda}=\bm{0}$, i.e., the added diagonal trick term in Eq. (\ref{eq:ext-GFEapp}) 
automatically vanishes when $G_{\mrm{app}}(\bm{m}) = G(\bm{m})$.

\bibliography{I-SusP}

\begin{thebibliography}{22}%
\makeatletter
\providecommand \@ifxundefined [1]{%
 \@ifx{#1\undefined}
}%
\providecommand \@ifnum [1]{%
 \ifnum #1\expandafter \@firstoftwo
 \else \expandafter \@secondoftwo
 \fi
}%
\providecommand \@ifx [1]{%
 \ifx #1\expandafter \@firstoftwo
 \else \expandafter \@secondoftwo
 \fi
}%
\providecommand \natexlab [1]{#1}%
\providecommand \enquote  [1]{``#1''}%
\providecommand \bibnamefont  [1]{#1}%
\providecommand \bibfnamefont [1]{#1}%
\providecommand \citenamefont [1]{#1}%
\providecommand \href@noop [0]{\@secondoftwo}%
\providecommand \href [0]{\begingroup \@sanitize@url \@href}%
\providecommand \@href[1]{\@@startlink{#1}\@@href}%
\providecommand \@@href[1]{\endgroup#1\@@endlink}%
\providecommand \@sanitize@url [0]{\catcode `\\12\catcode `\$12\catcode
  `\&12\catcode `\#12\catcode `\^12\catcode `\_12\catcode `\%12\relax}%
\providecommand \@@startlink[1]{}%
\providecommand \@@endlink[0]{}%
\providecommand \url  [0]{\begingroup\@sanitize@url \@url }%
\providecommand \@url [1]{\endgroup\@href {#1}{\urlprefix }}%
\providecommand \urlprefix  [0]{URL }%
\providecommand \Eprint [0]{\href }%
\providecommand \doibase [0]{http://dx.doi.org/}%
\providecommand \selectlanguage [0]{\@gobble}%
\providecommand \bibinfo  [0]{\@secondoftwo}%
\providecommand \bibfield  [0]{\@secondoftwo}%
\providecommand \translation [1]{[#1]}%
\providecommand \BibitemOpen [0]{}%
\providecommand \bibitemStop [0]{}%
\providecommand \bibitemNoStop [0]{.\EOS\space}%
\providecommand \EOS [0]{\spacefactor3000\relax}%
\providecommand \BibitemShut  [1]{\csname bibitem#1\endcsname}%
\let\auto@bib@innerbib\@empty
\bibitem [{\citenamefont {Opper}\ and\ \citenamefont
  {Saad(Eds.)}(2001)}]{Opper&Saad2001}%
  \BibitemOpen
  \bibfield  {author} {\bibinfo {author} {\bibfnamefont {M.}~\bibnamefont
  {Opper}}\ and\ \bibinfo {author} {\bibfnamefont {D.}~\bibnamefont
  {Saad(Eds.)}},\ }\href@noop {} {\emph {\bibinfo {title} {Advanced Mean Field
  Methods---Theory and Practice}}}\ (\bibinfo  {publisher} {MIT Press},\
  \bibinfo {year} {2001})\BibitemShut {NoStop}%
\bibitem [{\citenamefont {M{\'e}zard}\ and\ \citenamefont
  {Montanari}(2009)}]{Mezard&Montanari2009}%
  \BibitemOpen
  \bibfield  {author} {\bibinfo {author} {\bibfnamefont {M.}~\bibnamefont
  {M{\'e}zard}}\ and\ \bibinfo {author} {\bibfnamefont {A.}~\bibnamefont
  {Montanari}},\ }\href@noop {} {\emph {\bibinfo {title} {Information, Physics
  and Computation}}}\ (\bibinfo  {publisher} {Oxford University Press},\
  \bibinfo {year} {2009})\BibitemShut {NoStop}%
\bibitem [{\citenamefont {Sanchez}(1982)}]{Sanchez1982}%
  \BibitemOpen
  \bibfield  {author} {\bibinfo {author} {\bibfnamefont {J.~M.}\ \bibnamefont
  {Sanchez}},\ }\href@noop {} {\bibfield  {journal} {\bibinfo  {journal}
  {Physica A}\ }\textbf {\bibinfo {volume} {111}},\ \bibinfo {pages} {200}
  (\bibinfo {year} {1982})}\BibitemShut {NoStop}%
\bibitem [{\citenamefont {Morita}(1990)}]{Morita1990}%
  \BibitemOpen
  \bibfield  {author} {\bibinfo {author} {\bibfnamefont {T.}~\bibnamefont
  {Morita}},\ }\href@noop {} {\bibfield  {journal} {\bibinfo  {journal}
  {Progress of Theoretical Physics}\ }\textbf {\bibinfo {volume} {85}},\
  \bibinfo {pages} {243} (\bibinfo {year} {1990})}\BibitemShut {NoStop}%
\bibitem [{\citenamefont {Tanaka}(2003)}]{K.Tanaka2003}%
  \BibitemOpen
  \bibfield  {author} {\bibinfo {author} {\bibfnamefont {K.}~\bibnamefont
  {Tanaka}},\ }\href@noop {} {\bibfield  {journal} {\bibinfo  {journal} {IEICE
  Trans. on Information and Systems}\ }\textbf {\bibinfo {volume} {E86-D}},\
  \bibinfo {pages} {1228} (\bibinfo {year} {2003})}\BibitemShut {NoStop}%
\bibitem [{\citenamefont {Welling}\ and\ \citenamefont
  {Teh}(2003)}]{Welling&Teh2003}%
  \BibitemOpen
  \bibfield  {author} {\bibinfo {author} {\bibfnamefont {M.}~\bibnamefont
  {Welling}}\ and\ \bibinfo {author} {\bibfnamefont {Y.~W.}\ \bibnamefont
  {Teh}},\ }\href@noop {} {\bibfield  {journal} {\bibinfo  {journal}
  {Artificial Intelligence}\ }\textbf {\bibinfo {volume} {143}},\ \bibinfo
  {pages} {19} (\bibinfo {year} {2003})}\BibitemShut {NoStop}%
\bibitem [{\citenamefont {Welling}\ and\ \citenamefont
  {Teh}(2004)}]{Welling&Teh2004}%
  \BibitemOpen
  \bibfield  {author} {\bibinfo {author} {\bibfnamefont {M.}~\bibnamefont
  {Welling}}\ and\ \bibinfo {author} {\bibfnamefont {Y.~W.}\ \bibnamefont
  {Teh}},\ }\href@noop {} {\bibfield  {journal} {\bibinfo  {journal} {Neural
  Computation}\ }\textbf {\bibinfo {volume} {16}},\ \bibinfo {pages} {197}
  (\bibinfo {year} {2004})}\BibitemShut {NoStop}%
\bibitem [{\citenamefont {M{\'e}zard}\ and\ \citenamefont
  {Mora}(2009)}]{Mezard&Mora2009}%
  \BibitemOpen
  \bibfield  {author} {\bibinfo {author} {\bibfnamefont {M.}~\bibnamefont
  {M{\'e}zard}}\ and\ \bibinfo {author} {\bibfnamefont {T.}~\bibnamefont
  {Mora}},\ }\href@noop {} {\bibfield  {journal} {\bibinfo  {journal} {Journal
  of Physiology-Paris}\ }\textbf {\bibinfo {volume} {103}},\ \bibinfo {pages}
  {107} (\bibinfo {year} {2009})}\BibitemShut {NoStop}%
\bibitem [{\citenamefont {Pearl}(1988)}]{Pearl1988}%
  \BibitemOpen
  \bibfield  {author} {\bibinfo {author} {\bibfnamefont {J.}~\bibnamefont
  {Pearl}},\ }\href@noop {} {\emph {\bibinfo {title} {Probabilistic Reasoning
  in Intelligent Systems: Networks of Plausible Inference (2nd ed.)}}}\
  (\bibinfo  {publisher} {San Francisco, CA: Morgan Kaufmann},\ \bibinfo {year}
  {1988})\BibitemShut {NoStop}%
\bibitem [{\citenamefont {Bethe}(1935)}]{Bethe1935}%
  \BibitemOpen
  \bibfield  {author} {\bibinfo {author} {\bibfnamefont {H.~A.}\ \bibnamefont
  {Bethe}},\ }\href@noop {} {\bibfield  {journal} {\bibinfo  {journal}
  {Proceedings of the Royal Society of London. Series A: Mathematical and
  Physical Sciences}\ }\textbf {\bibinfo {volume} {150}},\ \bibinfo {pages}
  {552} (\bibinfo {year} {1935})}\BibitemShut {NoStop}%
\bibitem [{\citenamefont {Kabashima}\ and\ \citenamefont
  {Saad}(1998)}]{Kabashima&Saad1998}%
  \BibitemOpen
  \bibfield  {author} {\bibinfo {author} {\bibfnamefont {Y.}~\bibnamefont
  {Kabashima}}\ and\ \bibinfo {author} {\bibfnamefont {D.}~\bibnamefont
  {Saad}},\ }\href@noop {} {\bibfield  {journal} {\bibinfo  {journal}
  {Europhys. Lett.}\ }\textbf {\bibinfo {volume} {44}},\ \bibinfo {pages} {668}
  (\bibinfo {year} {1998})}\BibitemShut {NoStop}%
\bibitem [{\citenamefont {Yedidia}\ \emph {et~al.}(2001)\citenamefont
  {Yedidia}, \citenamefont {Freeman},\ and\ \citenamefont {Weiss}}]{GBP2001}%
  \BibitemOpen
  \bibfield  {author} {\bibinfo {author} {\bibfnamefont {J.~S.}\ \bibnamefont
  {Yedidia}}, \bibinfo {author} {\bibfnamefont {W.~T.}\ \bibnamefont
  {Freeman}}, \ and\ \bibinfo {author} {\bibfnamefont {Y.}~\bibnamefont
  {Weiss}},\ }\href@noop {} {\bibfield  {journal} {\bibinfo  {journal} {Neural
  Information Processing Systems (NIPS)}\ }\textbf {\bibinfo {volume} {13}},\
  \bibinfo {pages} {689} (\bibinfo {year} {2001})}\BibitemShut {NoStop}%
\bibitem [{\citenamefont {Yasuda}\ and\ \citenamefont
  {Tanaka}(2007)}]{Yasuda&Tanaka2007}%
  \BibitemOpen
  \bibfield  {author} {\bibinfo {author} {\bibfnamefont {M.}~\bibnamefont
  {Yasuda}}\ and\ \bibinfo {author} {\bibfnamefont {K.}~\bibnamefont
  {Tanaka}},\ }\href@noop {} {\bibfield  {journal} {\bibinfo  {journal} {J.
  Phys. A: Math. and Theor.}\ }\textbf {\bibinfo {volume} {40}},\ \bibinfo
  {pages} {9993} (\bibinfo {year} {2007})}\BibitemShut {NoStop}%
\bibitem [{\citenamefont {Kappen}\ and\ \citenamefont
  {Rodr{\'i}guez}(1998)}]{KR1998}%
  \BibitemOpen
  \bibfield  {author} {\bibinfo {author} {\bibfnamefont {H.~J.}\ \bibnamefont
  {Kappen}}\ and\ \bibinfo {author} {\bibfnamefont {F.~B.}\ \bibnamefont
  {Rodr{\'i}guez}},\ }\href@noop {} {\bibfield  {journal} {\bibinfo  {journal}
  {Neural Computation}\ }\textbf {\bibinfo {volume} {10}},\ \bibinfo {pages}
  {1137} (\bibinfo {year} {1998})}\BibitemShut {NoStop}%
\bibitem [{\citenamefont {Tanaka}(1998)}]{T.Tanaka1998}%
  \BibitemOpen
  \bibfield  {author} {\bibinfo {author} {\bibfnamefont {T.}~\bibnamefont
  {Tanaka}},\ }\href@noop {} {\bibfield  {journal} {\bibinfo  {journal} {Phys.
  Rev. E}\ }\textbf {\bibinfo {volume} {58}},\ \bibinfo {pages} {2302}
  (\bibinfo {year} {1998})}\BibitemShut {NoStop}%
\bibitem [{\citenamefont {Yasuda}\ and\ \citenamefont
  {Tanaka}(2009)}]{Yasuda&Tanaka2009}%
  \BibitemOpen
  \bibfield  {author} {\bibinfo {author} {\bibfnamefont {M.}~\bibnamefont
  {Yasuda}}\ and\ \bibinfo {author} {\bibfnamefont {K.}~\bibnamefont
  {Tanaka}},\ }\href@noop {} {\bibfield  {journal} {\bibinfo  {journal} {Neural
  Computation}\ }\textbf {\bibinfo {volume} {21}},\ \bibinfo {pages} {3130}
  (\bibinfo {year} {2009})}\BibitemShut {NoStop}%
\bibitem [{\citenamefont {Opper}\ and\ \citenamefont
  {Winther}(2001{\natexlab{a}})}]{Opper&Winther2001a}%
  \BibitemOpen
  \bibfield  {author} {\bibinfo {author} {\bibfnamefont {M.}~\bibnamefont
  {Opper}}\ and\ \bibinfo {author} {\bibfnamefont {O.}~\bibnamefont
  {Winther}},\ }\href@noop {} {\bibfield  {journal} {\bibinfo  {journal} {Phys.
  Rev. Lett.}\ }\textbf {\bibinfo {volume} {86}},\ \bibinfo {pages} {3695}
  (\bibinfo {year} {2001}{\natexlab{a}})}\BibitemShut {NoStop}%
\bibitem [{\citenamefont {Opper}\ and\ \citenamefont
  {Winther}(2001{\natexlab{b}})}]{Opper&Winther2001b}%
  \BibitemOpen
  \bibfield  {author} {\bibinfo {author} {\bibfnamefont {M.}~\bibnamefont
  {Opper}}\ and\ \bibinfo {author} {\bibfnamefont {O.}~\bibnamefont
  {Winther}},\ }\href@noop {} {\bibfield  {journal} {\bibinfo  {journal} {Phys.
  Rev. E}\ }\textbf {\bibinfo {volume} {64}},\ \bibinfo {pages} {056131}
  (\bibinfo {year} {2001}{\natexlab{b}})}\BibitemShut {NoStop}%
\bibitem [{\citenamefont {Marinari}\ and\ \citenamefont {{V. Van
  Kerrebroeck}}(2010)}]{Marinari&Kerrebroeck2010}%
  \BibitemOpen
  \bibfield  {author} {\bibinfo {author} {\bibfnamefont {E.}~\bibnamefont
  {Marinari}}\ and\ \bibinfo {author} {\bibnamefont {{V. Van Kerrebroeck}}},\
  }\href@noop {} {\bibfield  {journal} {\bibinfo  {journal} {J. Stat. Mech.:
  Theor. Exp.}\ }\textbf {\bibinfo {volume} {2010}},\ \bibinfo {pages} {P02008}
  (\bibinfo {year} {2010})}\BibitemShut {NoStop}%
\bibitem [{\citenamefont {Horiguchi}(1981)}]{Horiguchi1981}%
  \BibitemOpen
  \bibfield  {author} {\bibinfo {author} {\bibfnamefont {T.}~\bibnamefont
  {Horiguchi}},\ }\href@noop {} {\bibfield  {journal} {\bibinfo  {journal}
  {Physica A}\ }\textbf {\bibinfo {volume} {107}},\ \bibinfo {pages} {360}
  (\bibinfo {year} {1981})}\BibitemShut {NoStop}%
\bibitem [{\citenamefont {Yasuda}\ and\ \citenamefont
  {Tanaka}(2006)}]{Yasuda&Tanaka2006}%
  \BibitemOpen
  \bibfield  {author} {\bibinfo {author} {\bibfnamefont {M.}~\bibnamefont
  {Yasuda}}\ and\ \bibinfo {author} {\bibfnamefont {K.}~\bibnamefont
  {Tanaka}},\ }\href@noop {} {\bibfield  {journal} {\bibinfo  {journal} {J.
  Phys. Soc. Jpn.}\ }\textbf {\bibinfo {volume} {75}},\ \bibinfo {pages} {1}
  (\bibinfo {year} {2006})}\BibitemShut {NoStop}%
\bibitem [{Note1()}]{Note1}%
  \BibitemOpen
  \bibinfo {note} {This measure is generally not a distance in a precise
  mathematical sense, because $D(\protect \bm {A}||\protect \bm {B}) \not =
  D(\protect \bm {B}||\protect \bm {A})$.}\BibitemShut {Stop}%
\end{thebibliography}%
\end{document}